# 3-Qubit Circular Quantum Convolution Computation using Fourier Transform with Illustrative Examples

**Artyom M. Grigoryan**[1] **and Sos S. Agaian**[2]

[1] Department of Electrical and Computer Engineering, University of Texas at San Antonio, USA, email**:** amgrigoryan@utsa.edu, tel/fax: (210) 458-7518/5947

[2] Computer Science Department, The College of Staten Island New York, USA

**Abstract**
In this work, we describe examples for calculating the 1-D circular convolution of signals represented by 3-qubit superpositions. The case is considered, when the discrete Fourier transform of one of the signals is known and calculated in advance and only the QFT of another signal is calculated. The frequency characteristics of many linear time-invariant systems and filters are well known. Therefore, the considered method of convolution can be used for these systems in quantum computation. The ideal low pass and high pass filters are considered and quantum schemes for convolution are presented. The method of the Fourier transform is used with one addition qubit to prepare the quantum superposition for the inverse quantum Fourier transform.

## 1 Introduction

The fast Fourier transform-based convolution used in signal/image processing, data-driven learning, and deep convolutional neural networks numerous applications [1]-[11]. However, the direct application of the quantum Fourier transform (QFT) for calculation of the convolution faces many difficulties associated with finding gates for the multiplication of the Fourier transforms. Building an efficient quantum convolution algorithm for science and engineering needs is challenging. In quantum computation, the $r$-qubit quantum Fourier transform (QFT) is defined as the $2^r$-point discrete Fourier transform (DFT) of amplitudes of the quantum superposition of the signal of length $2^r$. Different algorithms and circuits for the QFT have been developed [12]-[16]. The direct application of the quantum Fourier transform for calculation of the convolution faces many difficulties, which are associated with finding gates for the multiplication of the Fourier transforms [18][19]. It should be noted that exists the opinion that the quantum convolution is "physically impossible" [20]. One solution here could be to use the multiplicative property, which states that the Fourier transform of a convolution of two signals is the pointwise product of their Fourier transforms under suitable conditions.

   In this paper, we propose a method of quantum convolution which is described in detail on the examples with 3 qubits. Here, we note that the convolution of length $2^r$, or $r$-qubit convolution, can be sequentially separated by short convolutions [17]. Therefore, the presence of the schemes for short convolutions will make probably it possible to implement the calculation of convolution on examples with signals of length $2^r, r \geq 3$. The case is considered, when the discrete Fourier transform of one of the signals is known and only the QFT of another signal is calculated. The considered method of convolution can be used for linear time-invariant systems with the known frequency characteristics. The Fourier transform method is very efficient when computing convolution in a classical computer; the convolution is reduced to multiplication. But it is this multiplication operation that is the most difficult step in quantum convolution using the Fourier transform. To overcome this obstacle, we suggest using an additional qubit and perform the corresponding permutation and prepare the quantum superposition of qubits for the inverse QFT. The examples of circuits for the low-pass and high-pass filters are also given.

## 2 Background

Quantum computing holds the promise of fast solutions to many problems in several areas, including quantum signal/image processing and quantum machine learning [29]. In recent decades, many papers have been published with the main goal of extending traditional signal and image processing tasks and operations to the framework of quantum computing. It is well known that efficient quantum algorithms exist and perform significantly faster than classical computers [21]. The basic concept in the quantum computation is the qubit described by the superposition of states $|q_1⟩ = a|0⟩ + b|1⟩$. This qubit may be in one of the basis states $|0⟩$ and $|1⟩$ with probability $p_0 = |a|^2$ and $p_1 = |b|^2$, respectively. Therefore, $|a|^2 + |b|^2 = 1$. The two-qubit superposition $|q_2⟩ = a|0⟩ + b|1⟩ + c|2⟩ + d|3⟩$ is for two qubits which may be in for basis states with probabilities $p_0 = |a|^2$, $p_1 = |b|^2$, $p_2 = |c|^2$, and $p_4 = |d|^2$. Thus, a single qubit is described by 2 classical bits, two qubits by 4 bits, and so on, $k$ qubits hold the same amount of information as $2^k$ bits.

All computation operations over qubits, or multiqubit superposition of states, are described by unitary matrices. This is a major hurdle in the construction of quantum circuits for many of the traditional operations that are widely used in signal and image processing. They include the convolution and gradient operators. In medical image processing, the method of quantum edge detection was described in [24]. Later, the model of image representation known as the novel enhanced quantum representation (NEQR) has proven to be very suitable for extracting edges with Sobel gradients [25][26]. We also mention the quantum algorithm for the Kirsch and Prewitt operator-based edge extractions [27][28]. The computation of linear and cyclic convolution in quantum computation is still the open problem. If the traditional fast method for convoluting two signals is based on the DFT and reduced to the pointwise multiplication of these transforms, such multiplication in quantum computations should be performed, or at least approximated, by unitary transforms. Thus, the circuits for the QFT exist [12]-[16] but cannot directly be applied for calculating the convolution [18][19]. Our inability to work with the QFT in the traditional way should lead us to developing additional methods and circuits that would solve this difficult problem. We believe that sooner or later this problem will be solved for many cases of the convolution.

## 3 Quantum Convolution

In this section, we describe the quantum scheme for the convolution of a signal $f_n$, $n = (N − 1)$, in a linear time-invariant system (LTI), when its frequency characteristic $H_p$, $p = (N − 1)$, is given. The length is a power of two, $N = 2^r, r > 1$. The $N$-point DFT of the signal is

$$F_p = \sum_{n=0}^{N-1} f_n W^{np}, \quad p = 0:(N-1). \tag{1}$$

Here, $W$ is the exponential coefficient $W_N = \exp(-i2\pi/N)$. For simplicity of calculation and drawing, we consider the $N = 8$ case. For the quantum superposition of the given signal $f_n$,

$$|\check{f}⟩ = \sum_{n=0}^{7} f_n |n⟩, \tag{2}$$

the quantum Fourier transform is described by the following 3-qubit superposition:

$$|\psi_3(\check{f})⟩ = \frac{1}{\sqrt{8}} \sum_{p=0}^{7} F_P |p⟩, \tag{3}$$

Here, $|n⟩$ and $|p⟩$ denote the basis states. We consider that the signal $f_n$ was normalized, i.e.,

$$\sum_{n=0}^{7} |f_n|^2 = 1. \tag{4}$$

The quantum algorithms for the Fourier transform are known [12]-[14] and the circuit unit element for the 3-qubit QFT can be presented, as shown in Fig. 1.

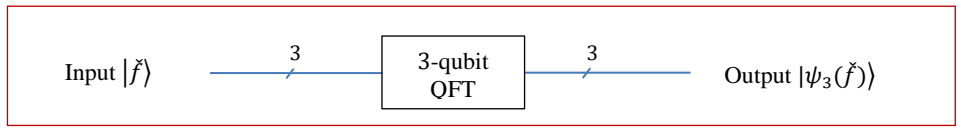

**Figure 1.** The circuit element for the 3-qubit QFT.

We take the amplitudes of the transform in the order $F_7, F_3, F_5, F_1, F_6, F_2, F_4$, and $F_0$, as they are calculated in the paired algorithm of the 3-qubit QFT [12]. After a permutation, this transform can be written in the order $F_7, F_1, F_6, F_2, F_5, F_3, F_4$, and $F_0$. The corresponding 3-qubit superposition is

$$|\psi_3(\check{f})\rangle = \frac{1}{\sqrt{8}}[(F_7|0\rangle + F_1|1\rangle) + (F_6|2\rangle + F_2|3\rangle) + (F_5|4\rangle + F_3|5\rangle) + (F_4|6\rangle + F_0|7\rangle)]. \quad (5)$$

The values of the frequency characteristic $H_p$ of the system are considered in the same order. These two sequences of numbers should be multiplied pointwise, to obtain the Fourier transform of the convolution

$$Y_p = F_p H_p, \quad p = 7,1,6,2,5,3,4,0.$$

Now, we consider the scheme given in Fig. 2. This scheme is for the state-wise multiplication of two qubits in matrix form,

$$|\check{a}\rangle = a_0|0\rangle + a_1|1\rangle \text{ and } |\check{b}\rangle = b_0|0\rangle + b_1|1\rangle,$$

by using the operation $U$. Here, $a_0^2 + a_1^2 = b_0^2 + b_1^2 = 1$, or $|a_0|^2 + |a_1|^2 = |b_0|^2 + |b_1|^2 = 1$, when the amplitudes are complex numbers.

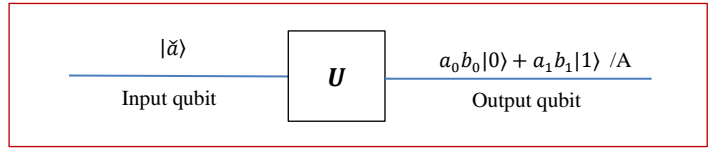

**Figure 2.** The abstract scheme for the operation $U$, ($A = \sqrt{|a_0 b_0|^2 + |a_1 b_1|^2}$).

After normalizing the superposition of qubits, the matrix representation of this operation can be written as

$$U|\check{a}\rangle = \begin{bmatrix} b_0 & 0 \\ 0 & b_1 \end{bmatrix} \begin{bmatrix} a_0 \\ a_1 \end{bmatrix} = \begin{bmatrix} b_0 a_0 \\ b_1 a_1 \end{bmatrix} = a_0 b_0 |0\rangle + a_1 b_1 |1\rangle \rightarrow \frac{1}{A}(a_0 b_0 |0\rangle + a_1 b_1 |1\rangle), \quad (6)$$

where the normalized coefficient for this new qubit is $A = \sqrt{|a_0 b_0|^2 + |a_1 b_1|^2}$. Thus, $U = U_{b_1, b_2}$ is the one-qubit operation. It can be considered as a Hermitian operator corresponding to measurement of the qubit in the standard basis [21]. We consider the case when the numbers $b_0$ and $b_1$ are known and not equal to zero; there is no need to measure the qubit $|\check{b}\rangle$.

The diagonal matrix $U$ can be considered with the determinant 1 as

$$U = U_{b_1, b_2} = \frac{1}{\sqrt{b_0 b_1}} \begin{bmatrix} b_0 & 0 \\ 0 & b_1 \end{bmatrix} = \frac{1}{\sqrt{\sin(2\varphi)/2}} \begin{bmatrix} \cos(\varphi) & 0 \\ 0 & \sin(\varphi) \end{bmatrix}, \quad (7)$$

for an angle $\varphi \in [0, 2\pi)$.

The following four matrices are defined:

$$U_0 = \frac{1}{\sqrt{H_4 H_0}} \begin{bmatrix} H_4 & 0 \\ 0 & H_0 \end{bmatrix}, \quad U_3 = \frac{1}{\sqrt{H_5 H_3}} \begin{bmatrix} H_5 & 0 \\ 0 & H_3 \end{bmatrix}, \quad U_2 = \frac{1}{\sqrt{H_6 H_2}} \begin{bmatrix} H_6 & 0 \\ 0 & H_2 \end{bmatrix},$$

$$U_1 = \frac{1}{\sqrt{H_7 H_1}} \begin{bmatrix} H_7 & 0 \\ 0 & H_1 \end{bmatrix}.$$

It is assumed that all numbers $H_p \neq 0$. The case when one of them is equal to zero should be considered separately. The scheme for multiplication of two transforms, $Y_p = F_p H_p$, by using these four matrices is shown in Fig. 3. The first two bits are used as control bits to apply one of these matrices. The matrix $U_0$ is applied when the first two bits are 1. The matrix $U_3$ is applied when the first two bits are 1 and 0, respectively. The matrix $U_2$ is applied when these two bits are 0 and 1, and the matrix $U_1$ is applied when the first two bits are 0.

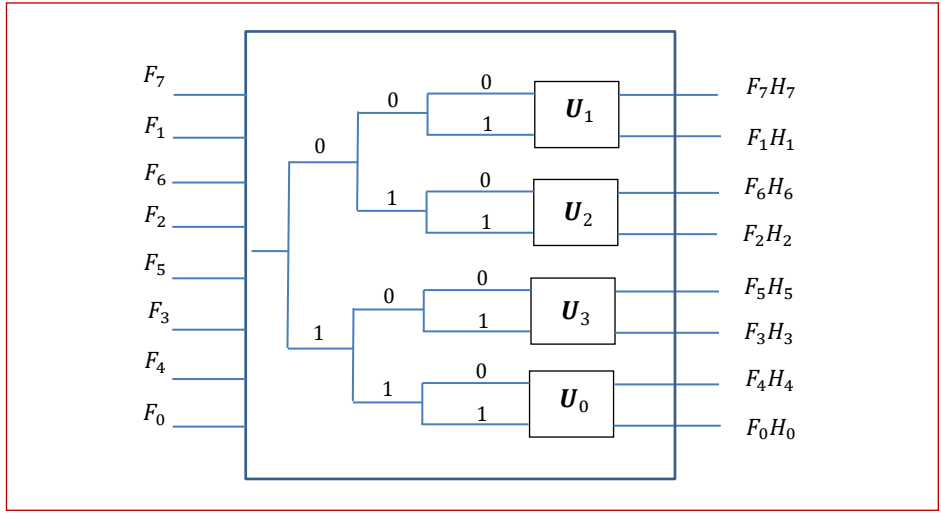

**Figure 3.** The scheme for the multiplication of transforms.

In quantum computation, the matrices on one qubit operators are unitary. Therefore, we consider the case when the signal $f_n$ is convoluted with a real impulse response $h_n$ of the filter or system. Then, there are three complex conjugate pairs, $H_7 = \overline{H}_1$, $H_6 = \overline{H}_2$, and $H_5 = \overline{H}_3$. Denoting by $\varphi_1$, $\varphi_2$, and $\varphi_3$ the phases of the numbers $H_1, H_2,$ and $H_3$, respectively, we obtain the following unitary matrices:

$$U_3 = \begin{bmatrix} e^{-i\varphi_3} & 0 \\ 0 & e^{i\varphi_3} \end{bmatrix}, \qquad U_2 = \begin{bmatrix} e^{-i\varphi_2} & 0 \\ 0 & e^{i\varphi_2} \end{bmatrix}, \qquad U_1 = \begin{bmatrix} e^{-i\varphi_1} & 0 \\ 0 & e^{i\varphi_1} \end{bmatrix}. \qquad (8)$$

The numbers $H_4$ and $H_0$ are real, therefore the matrix $U_0$ is considered to be the 2×2 identity matrix, $I_2$. Namely, it is the diagonal matrix with coefficients $\pm 1$, since $H_4$ and $H_0$ can be negative. The matrices $U_1$, $U_2$, and $U_3$ correspond to the phase rotations by the angles $\varphi_1$, $\varphi_2$, and $\varphi_3$, respectively.

The scheme of multiplication of the transform $F_p$ by the phase coefficients of the frequency characteristic $H_p$ is shown in Fig. 4. Thus, this diagram is for the calculation of the transform $F_p H_p / |H_p|$.

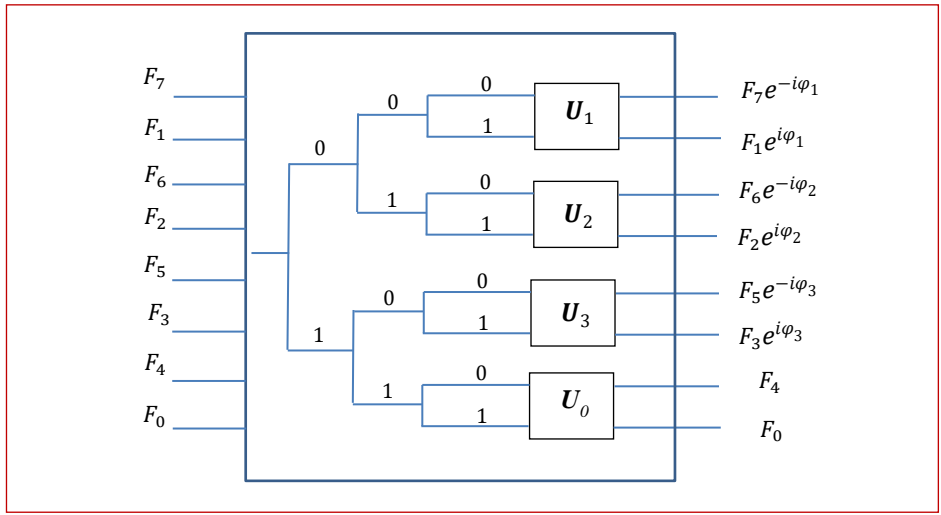

**Figure 4.** The scheme for the multiplication of the transform by phase coefficients.

The same diagram as the quantum circuit is shown in Fig. 5. The first two qubits are the control qubits. A bullet in the line indicates that the control qubit is in state $|1\rangle$, and an open circle indicates that the control qubit is in state $|0\rangle$, when applying the operators $U_k$, $k = 0{:}3$. One can see parallelism in computation, which is the hallmark of quantum computing.

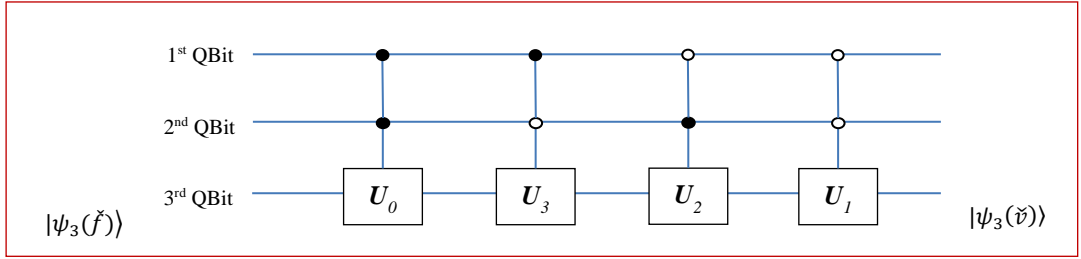

**Figure 5.** The quantum circuit for the multiplication of transform by the phase coefficients.

The 3-qubit superposition that corresponds to the multiplication of the Fourier transform by the phase coefficients can be written as

$$|\psi_3(\check{v})\rangle = \frac{1}{\sqrt{8}}[(F_7 e^{-i\varphi_1}|0\rangle + F_1 e^{i\varphi_1}|1\rangle) + (F_6 e^{-i\varphi_2}|2\rangle + F_2 e^{i\varphi_2}|3\rangle) + (F_5 e^{-i\varphi_3}|4\rangle + F_3 e^{i\varphi_3}|5\rangle) + (F_4|6\rangle + F_0|7\rangle)].$$

***Example 1:*** If the impulse response is $h = [1\ 1\ 0\ 0\ 0\ 0\ 0\ 0]/2$, the discrete Fourier transform equals

$$H_1 = 0.8536 - 0.3536i = 0.9239 e^{-i0.3927},$$
$$H_2 = 0.5 - 0.5i = 0.7071 e^{-i0.7854},$$
$$H_3 = 0.1464 - 0.3536i = 0.3827 e^{-i1.1781},$$
$$H_4 = 0, \quad H_0 = 1.$$

The values of phases are $\varphi_1 = -0.3927$, $\varphi_2 = -0.7854$, and $\varphi_3 = -1.1781$.

To get the values $Y_p$ of the Fourier transform of the convolution $y_n$, the amplitudes of the superposition $|\psi_3(\check{v})\rangle$ should be multiplied by magnitudes of the transfer function, as shown in Fig. 6.

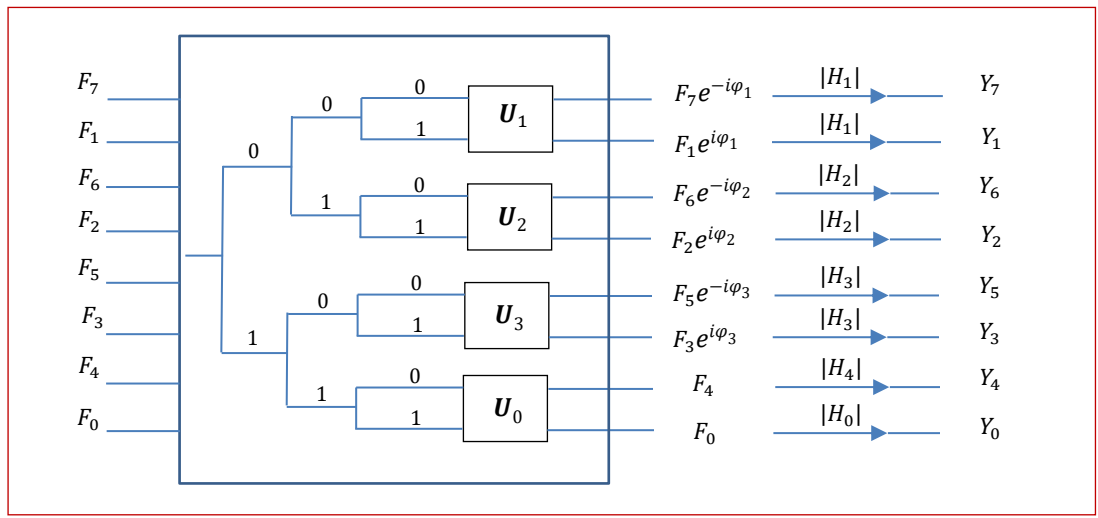

**Figure 6.** The abstract scheme for multiplication of the transforms.

This operation is described by the diagonal matrix

$$\boldsymbol{D} = \text{diag}\{|H_1|, |H_1|, |H_2|, |H_2|, |H_3|, |H_3|, |H_4|, |H_0|\}$$
$$= (|H_1|\boldsymbol{I}_2) \oplus (|H_2|\boldsymbol{I}_2) \oplus (|H_3|\boldsymbol{I}_2) \oplus \text{diag}\{|H_4|, |H_0|\} \quad (9)$$

For the above example, the diagonal matrix is

$$\boldsymbol{D} = \text{diag}\{0.9239, 0.9239, 0.7071, 0.7071, 0.3827, 0.3827, 0, 1\}.$$

If the frequency characteristic has zeros, then we can consider adding a constant, $H'_p = H_p + \text{const}$. For instance, for the above example, we can take $H'_p = H_p + 1 \neq 0$, $p = 0\colon 7$. The impulse response function will be changed as $h'_n = h_n + \delta_n$, i.e., the unit impulse will be added. The convolution changes as $y'_n = y_n + f_n$. Therefore, the original convolution can be calculated as $y_n = y'_n - f_n$, after the measurement the convolution $y'_n$.

The same control bits cannot be used for the multiplication by magnitudes $|H_p|$ in the way as done for the scheme with the phase operators in Fig. 4. One-qubit operation

$$a|0\rangle + b|1\rangle \rightarrow |H_1|a|0\rangle + |H_1|b|1\rangle$$

is the identity transformation. It does not change the qubit, since the amplitudes of the basis states should be normalized. In other words, $|H_1|a|0\rangle + |H_1|b|1\rangle = a|0\rangle + b|1\rangle$, for any value of $|H_1| \neq 0$. So we can think about rearranging the outputs, in order to perform the multiplication by the magnitudes $|H_p|$ of the transfer function. For example, we consider the permutation (**P**) with the natural order of outputs, which is shown in Fig. 7,

$$|\psi_3(\check{v})\rangle = \frac{1}{\sqrt{8}}[F_0|0\rangle + F_1 e^{i\varphi_1}|1\rangle + F_2 e^{i\varphi_2}|2\rangle + F_3 e^{i\varphi_3}|3\rangle + F_4|4\rangle + F_5 e^{-i\varphi_3}|5\rangle + F_6 e^{-i\varphi_2}|6\rangle$$
$$+ F_7 e^{-i\varphi_1}|7\rangle].$$

We denote the amplitudes of this superposition by $V_p$ and write

$$|\psi_3(\check{v})\rangle = \frac{1}{\sqrt{8}} \sum_{p=0}^{7} V_p |p\rangle. \quad (10)$$

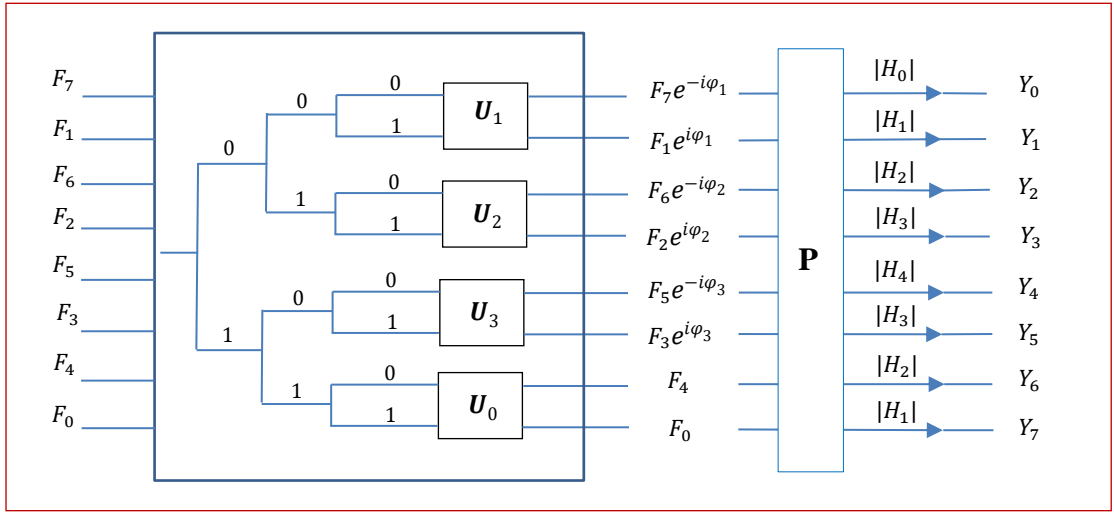

**Figure 7.** The abstract scheme for multiplication of the transforms with the permutation (P).

We denote by $\boldsymbol{D}$ the operator with the diagonal matrix and add it to the circuit as shown in Fig. 8.

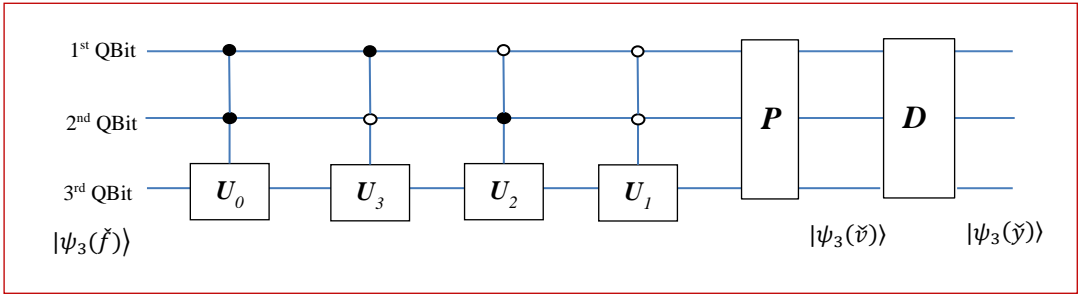

**Figure 8.** The abstract quantum circuit for the multiplication of the Fourier transforms.

This operator can be viewed as the process of preparing the next quantum superposition from $|\psi_3(\check{v})\rangle$,

$$|\psi_3(\check{y})\rangle = \frac{1}{A}[|H_0|V_0|0\rangle + |H_1|V_1|1\rangle + |H_2|V_2|2\rangle + |H_2|V_3|3\rangle + |H_4|V_5|4\rangle + |H_3|V_5|5\rangle + |H_2|V_6|6\rangle + \\ + |H_1|V_7|7\rangle]. \qquad (11)$$

The normalized coefficient equals

$$A = \sqrt{\sum_{p=0}^{7}|Y_p|^2} = \sqrt{8\sum_{n=0}^{7}|y_n|^2}.$$

Thus, $D$ is an operator of transition from one 3-qubit superposition to another. The amplitudes $V_p$, $p = 0:7$, are calculated by the quantum scheme in Fig. 5. The amplitudes of the transfer function $|H_p|$ are given. For the convolution in Example 1, the superposition for the convolution in frequency domain equals

$$|\psi_3(\check{y})\rangle = \frac{1}{A}[V_0|0\rangle + 0.9239(V_1|1\rangle + V_7|7\rangle) + 0.7071(V_2|2\rangle + V_6|6\rangle) + 0.3827(V_3|3\rangle + V_5|5\rangle)].$$

In quantum computation, this step refers to the measurement. In other words, the operator $D$ is the Hermitian operator specifying the measurement in a 3-qubit system in the standard basis,

$$D = |H_0||0\rangle\langle 0| + |H_1||1\rangle\langle 1| + |H_2||2\rangle\langle 2| + |H_3||3\rangle\langle 3| + |H_4||4\rangle\langle 4| + |H_3||5\rangle\langle 5| + |H_6||6\rangle\langle 6| \\ + |H_1||7\rangle\langle 7|.$$

The above superposition $|\psi_3(\check{y})\rangle$ will be the input for the inverse 3-qubit QFT.

The above circuit of multiplication of the transforms with the measurements in Fig. 8 can be considered as a circuit element shown in Fig. 9.

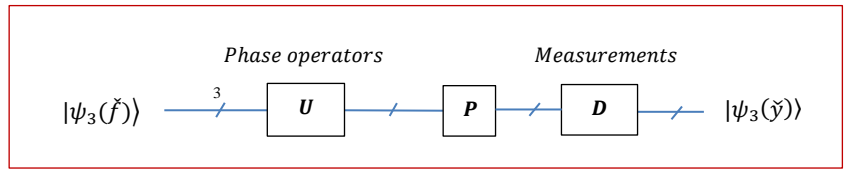

**Figure 9.** The abstract circuit element for the 3-qubit multiplication of the Fourier transform.

The complete quantum scheme for calculating the 3-qubit convolution with the inverse QDT is shown in Fig. 10.

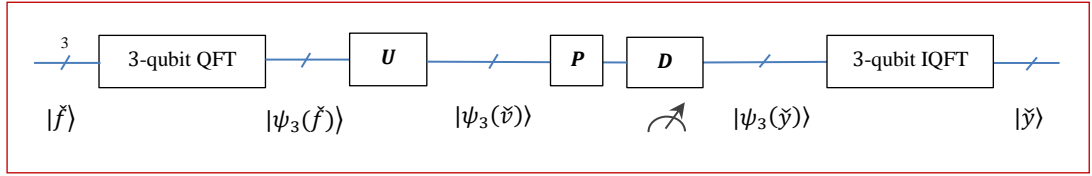

**Figure 10.** The scheme for computing the 3-qubit convolution of the signal.

It should be noted that the operation $D$ can be placed before the phase operators, as shown in Fig. 11. Then, the superpositions

$$|\psi_3(\check{v})\rangle = \frac{1}{A}\sum_{p=0}^{7} V_p|p\rangle = \frac{1}{A}\sum_{p=0}^{7} F_p|H_p||p\rangle$$

and

$$|\psi_3(\check{y})\rangle = \frac{1}{A}[V_0|0\rangle + V_1 e^{i\varphi_1}|1\rangle + V_2 e^{i\varphi_2}|2\rangle + V_3 e^{i\varphi_3}|3\rangle + V_4|4\rangle + V_5 e^{-i\varphi_3}|5\rangle + V_6 e^{-i\varphi_2}|6\rangle + V_7 e^{-i\varphi_1}|7\rangle].$$

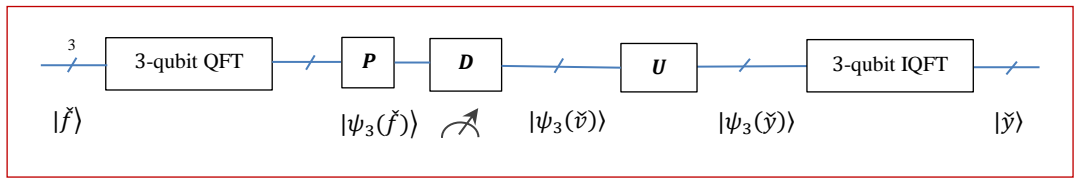

**Figure 11.** The scheme for computing the 3-qubit convolution of the signal.

It is important to note that the operator $D$ is not unitary. It could be added to the set of unitary operators in quantum computation, in order to use the operator $D$ in quantum circuits. The example below shows that it is possible to implement this operator without measurement and therefore solve the problem of quantum convolution at least for simple filters. Such filters are the low-pass and high-pass, and the band-pass ideal filters.

*Example 2:* Consider the following ideal low-pass filter

$$H_p = \begin{cases} 1, & p = 0, \\ e^{i\pi/12}, & p = 1, \\ e^{-i\pi/12}, & p = 7, \\ 0, & p = 2,3,4,5,6. \end{cases}$$

The impulse response $h = (2.9319, 2, 0.4824, -0.7321, -0.9319, 0, 1.5176, 2.7321)/8$ of this filter is shown in Fig. 12.

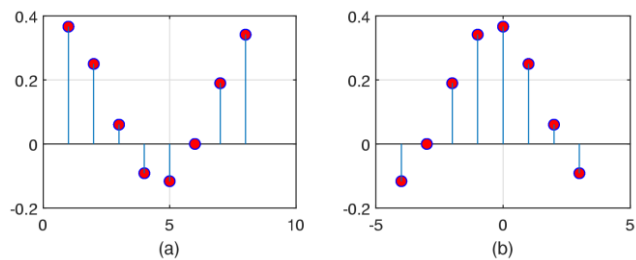

**Figure 12.** (a) The impulse response and (b) periodically shifted to the center.

The discrete Fourier transform of the convolution $y_n = f_n * h_n$ consists of only three components of the transform $F_p$, as shown in Fig. 13. The quantum Fourier transform of the convolution can be written as the following 3-qubit superposition:

$$|\psi_3(\breve{y})\rangle = \frac{1}{A}(Y_0|000\rangle + Y_1|001\rangle + Y_7|111\rangle) = \frac{1}{A}(H_0 F_0|000\rangle + H_1 F_1|001\rangle + H_7 F_7|111\rangle).$$

Thus, the input of the inverse 3-qubit QFT is the 3-qubit superposition

$$|\psi_3(\breve{y})\rangle = \frac{1}{A}\left(F_0|000\rangle + e^{i\pi/12} F_1|001\rangle + e^{-i\pi/12} F_7|111\rangle\right) \tag{12}$$

with the normalized coefficient $A = \sqrt{|F_0|^2 + |F_1|^2 + |F_7|^2}$. The matrix of phase operator

$$\boldsymbol{U}_1 = \begin{bmatrix} e^{-i\pi/12} & 0 \\ 0 & e^{i\pi/12} \end{bmatrix}.$$

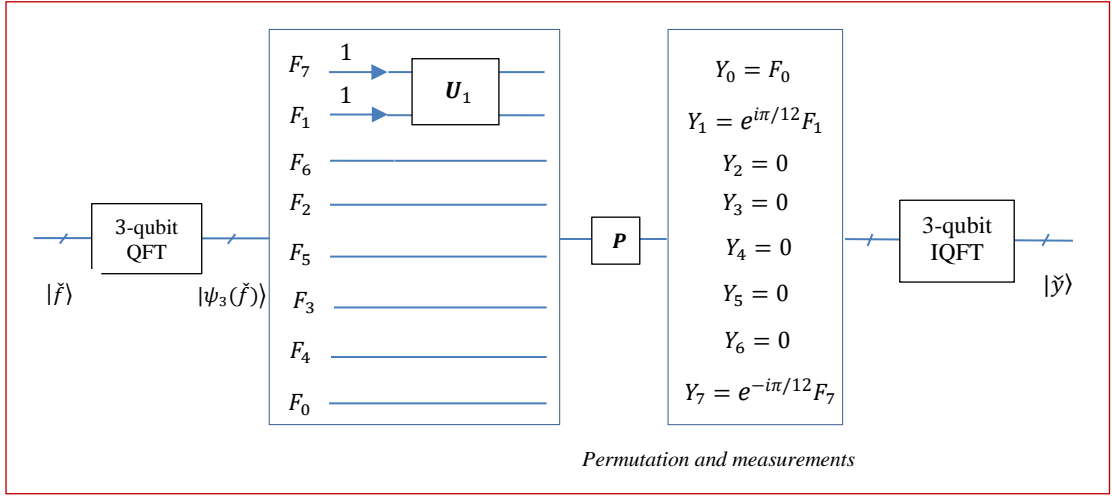

**Figure 13.** The scheme of calculation of the 8-point circular convolution.

The only difficulty lies in understanding how to implement quantum gates for the superposition in Eq. 12 after computing the 3-qubit QFT of the signal $f_n$. Five amplitudes of the 3-qubit superposition are zeroed. To prepare the superposition $|\psi_3(\breve{y})\rangle$, three qubits should be measured in the basis states $|000\rangle$, $|001\rangle$, and $|111\rangle$. Since $|H_0| = |H_1| = |H_7| = 1$, the operator $\boldsymbol{D}$ can be considered the Hermitian operator specifying the measurement in a 3-qubit system in the standard basis

$$\boldsymbol{D} = |0\rangle\langle 0| + |1\rangle\langle 1| + |H_1||7\rangle\langle 7|$$
$$= |000\rangle\langle 000| + |001\rangle\langle 001| + |111\rangle\langle 111| \tag{13}$$

with the matrix representation $\boldsymbol{D} = \mathrm{diag}\{1,1,0,0,0,0,0,1\}$.

Let us see how the diagonal operator $\boldsymbol{D}$ can be included in a non-abstract quantum scheme in this example. We add one qubit, qubit number 4, as the zero qubit. In other words, we consider the 4-qubit register, wherein the second part is filled by zeros. This superposition is

$$|\psi_4(\breve{y})\rangle = \begin{bmatrix} F_0 \\ e^{\frac{i\pi}{12}} F_1 \\ F_6 \\ F_2 \\ F_5 \\ F_3 \\ F_4 \\ e^{-\frac{i\pi}{12}} F_7 \end{bmatrix} \oplus \begin{bmatrix} 0 \\ 0 \\ 0 \\ 0 \\ 0 \\ 0 \\ 0 \\ 0 \end{bmatrix}. \tag{14}$$

Now, we consider a permutation that removes unnecessary information from the first part of the register for further processing. For instance, we consider the following permutation:

$$P_4: \begin{bmatrix} F_0 \\ e^{\frac{i\pi}{12}}F_1 \\ F_6 \\ F_2 \\ F_5 \\ F_3 \\ F_4 \\ e^{-\frac{i\pi}{12}}F_7 \end{bmatrix} \oplus \begin{bmatrix} 0 \\ 0 \\ 0 \\ 0 \\ 0 \\ 0 \\ 0 \\ 0 \end{bmatrix} \rightarrow |\psi_4(p)\rangle = \begin{bmatrix} F_0 \\ e^{\frac{i\pi}{12}}F_1 \\ 0 \\ 0 \\ 0 \\ 0 \\ 0 \\ e^{-\frac{i\pi}{12}}F_7 \end{bmatrix} \oplus \begin{bmatrix} 0 \\ 0 \\ F_2 \\ F_3 \\ F_4 \\ F_5 \\ F_6 \\ 0 \end{bmatrix}. \quad (15)$$

There are many permutations that result in the 16-dimensional vector with the first part equal $(F_0, e^{-i\pi/12}F_1, 0,0,0,0,0, e^{-i\pi/12}F_7)$. It is also known, that if an operator is unitary, it can be implemented in a quantum computer [22][23]. Thus, we consider that $P_4$ is a permutation that has a quantum circuit.

One more step is required to perform the 3-qubit inverse quantum Fourier transform on the first part of the 4-qubit register/superposition. For that, the first qubit can be used as the control qubit. The 3-qubit IQFT is performed when the first qubit is in the state 0. As a result, we obtain the circuit with the 4-qubit input, to calculate the 3-qubit convolution. This simplified circuit is shown in Fig. 14. The 3-qubit direct QFT and inverse QFT (IQFT) work in this circuit when the first qubit is in the state $|0\rangle$.

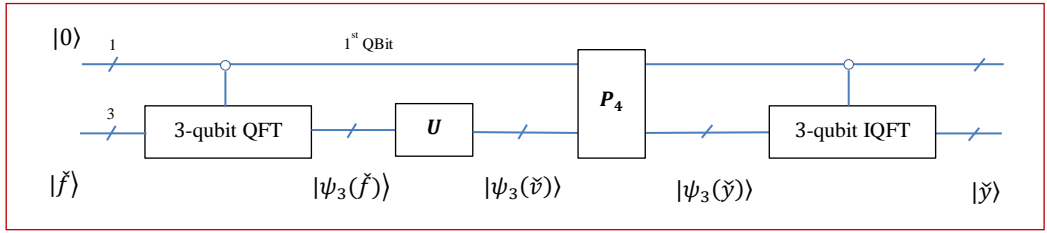

**Figure 14.** The quantum circuit for computing the 3-qubit convolution of the signal.

If use the 3-qubit inverse QFT (IQFT) when the first qubit is in the state $|1\rangle$, the result will be convolution of the signal with the high-pass filter

$$H_p = \begin{cases} 0, & p = 0,1,7, \\ 1, & p = 2,3,4,5,6. \end{cases}$$

We can simplify the circuit for this high-pass filter, as shown in Fig. 15. There is no need to use the phase operator $U$, and the permutation operator can be defined as

$$P_4: (|0\rangle + |1\rangle) \otimes \sum_{p=0}^{7} F_p |p\rangle = \begin{bmatrix} F_0 \\ F_1 \\ F_6 \\ F_2 \\ F_5 \\ F_3 \\ F_4 \\ F_7 \end{bmatrix} \oplus \begin{bmatrix} 0 \\ 0 \\ 0 \\ 0 \\ 0 \\ 0 \\ 0 \\ 0 \end{bmatrix} \rightarrow |\psi_4(p)\rangle = \begin{bmatrix} 0 \\ 0 \\ F_2 \\ F_3 \\ F_4 \\ F_5 \\ F_6 \\ 0 \end{bmatrix} \oplus \begin{bmatrix} F_0 \\ F_1 \\ 0 \\ 0 \\ 0 \\ 0 \\ 0 \\ F_7 \end{bmatrix}. \quad (16)$$

We assume that such examples of filtration or quantum convolution over $r$ qubits in the general case $r>3$ can be found, too. We can also go the other way in calculating the quantum convolution. The discrete paired transform allows to reduce the $2^r$-point cyclic convolution to the convolutions of lengths $2^{r-1}, 2^{r-2}, \dots, 8, 4, 2$, and 1 [17]. This process can be continued until we get all cyclic convolutions of the order less than or equal to 8. The short convolutions, namely the 2-point and 4-point cyclic convolutions, or 1- and 2-qubit convolutions, have simple schemes of calculation. Indeed, the 1-qubit convolution is described by the unitary matrix

$$U = \frac{1}{\sqrt{h_0^2 + h_1^2}} \begin{bmatrix} h_0 & h_1 \\ h_1 & h_0 \end{bmatrix}. \quad (13)$$

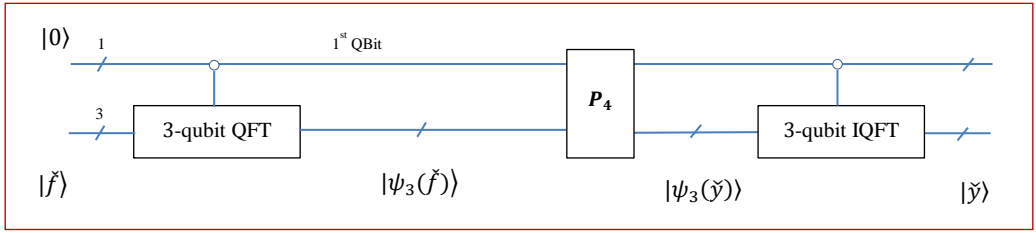

**Figure 15.** The quantum circuit for computing the 3-qubit convolution of the signal for the high-pass filter.

The scheme for 2-qubit convolution can be described by the operations like the above $r = 3$ case. For instance for the real impulse response, the multiplication of the Fourier transform $\{F_3, F_1, F_2, F_0\}$ by the phase coefficients and diagonal matrix $D = \text{diag}\{|H_0|, |H_1|, |H_2|, |H_1|\}$ can be performed by the scheme shown in Fig. 16. This matrix represents the Hermitian operator of measurements in the 2-qubit system,

$$D = |H_0||00\rangle\langle 00| + |H_1||01\rangle\langle 01| + |H_2||10\rangle\langle 10| + |H_1||11\rangle\langle 11|.$$

In this circuit, the permutation is $P_2 = (0,3)$, i.e., $P_2: (0,1,2,3) \rightarrow (3,1,2,0)$, and its matrix is

$$P_2 = \begin{bmatrix} 0 & 0 & 0 & 1 \\ 0 & 1 & 0 & 0 \\ 0 & 0 & 1 & 0 \\ 1 & 0 & 0 & 0 \end{bmatrix} \quad \text{and} \quad P_2 = (P_2)^{-1}.$$

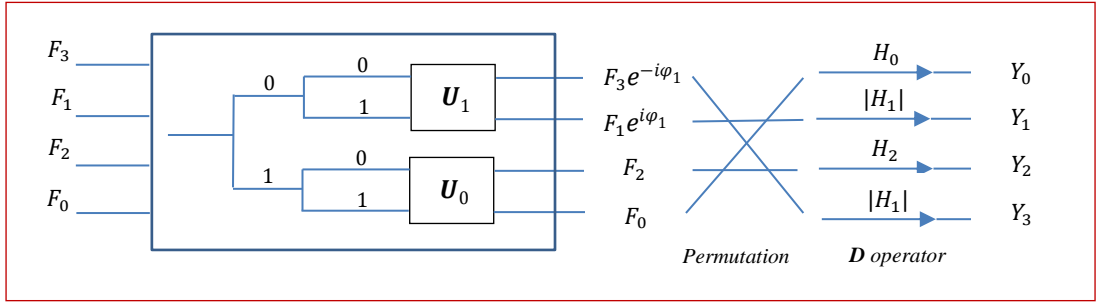

**Figure 16.** The scheme for multiplication of the 2-qubit transforms with the permutation.

## CONCLUSION

Implementing the circular convolution in quantum computers is difficult, but examples of convolution can be found. In this work, the convolution of a 3-qubit signal with another one is described. The second signal is the impulse response of a linear time-invariant system, whose discrete Fourier transform is known or given. The abstract quantum scheme for the 3-qubit circular convolution is proposed. The example with a low-pass filter is described. The most interesting case is when the impulse response of the system is real. The only difficulty in implementing this algorithm in the general case is the realization of operation **D** on qubits, i.e., the amplification of amplitudes of the 3-qubit superposition by absolute values of the system's frequency characteristic. For that, we propose to use one additional qubit. Measurements of qubits can be used for this operation.